\documentclass[pra,twocolumn,superscriptaddress,showpacs]{revtex4}
\usepackage{graphicx}
\usepackage{amsmath}
\usepackage{color}
\usepackage{array}
\usepackage{color}

\newcommand\ex{\mathrm{ex}}
\newcommand\Tr{\mathrm{Tr}}
\newcommand\ph{\mathrm{ph}}

\newcommand{\tabincell}[2]{\begin{tabular}{@{}#1@{}}#2\end{tabular}}

\begin{document}

\title{How bilayer excitons can greatly enhance thermoelectric efficiency}
\author{Kai Wu}
\email{kaiwu@stanford.edu}
\affiliation{Stanford Institute for Materials and Energy Sciences,
SLAC National Accelerator Laboratory, 2575 Sand Hill Road, Menlo Park, CA 94025, USA} 
\author{Louk Rademaker}
\affiliation{Kavli Institute for Theoretical Physics, University of California Santa Barbara, CA 93106, USA}

\author{Jan Zaanen}
\affiliation{Institute-Lorentz for Theoretical Physics, Leiden University, PO Box 9506, Leiden, The Netherlands}

\begin{abstract}
Presently, a major nanotechnological challenge is to design thermoelectric devices that have a high figure of merit. To that end, we propose to use bilayer excitons in two-dimensional nanostructures. Bilayer exciton systems are shown to have an improved thermopower and an enhanced electric counterflow and thermal conductivity, with respect to regular semiconductor-based thermoelectrics. We suggest an experimental realization of a bilayer exciton thermocouple. Based on current experimental parameters, a bilayer exciton heterostructures of $p$- and $n$-doped Bi$_2$Te$_3$ can have a figure of merit $zT \sim 60$. Another material suggestion is to make a bilayer out of electron-doped SrTiO$_3$ and hole-doped Ca$_3$Co$_4$O$_9$.
\end{abstract}


\maketitle

Ranging from household refrigeration to waste heat in power plants, the problem of converting heat into electricity or vice versa is of paramount technological importance. Instead of solving this issue with compression-based heat pumps and the like, solid state materials might offer a resolution. Among the notable advantages of using solid state Peltier or Seebeck devices are their light weight, the absence of any moving parts, and the elimination of any environmentally unfriendly substances. However, the main problem remains to devise a thermoelectric device with sufficient efficiency.

\begin{figure}
\begin{center}
	\includegraphics[width=\columnwidth]{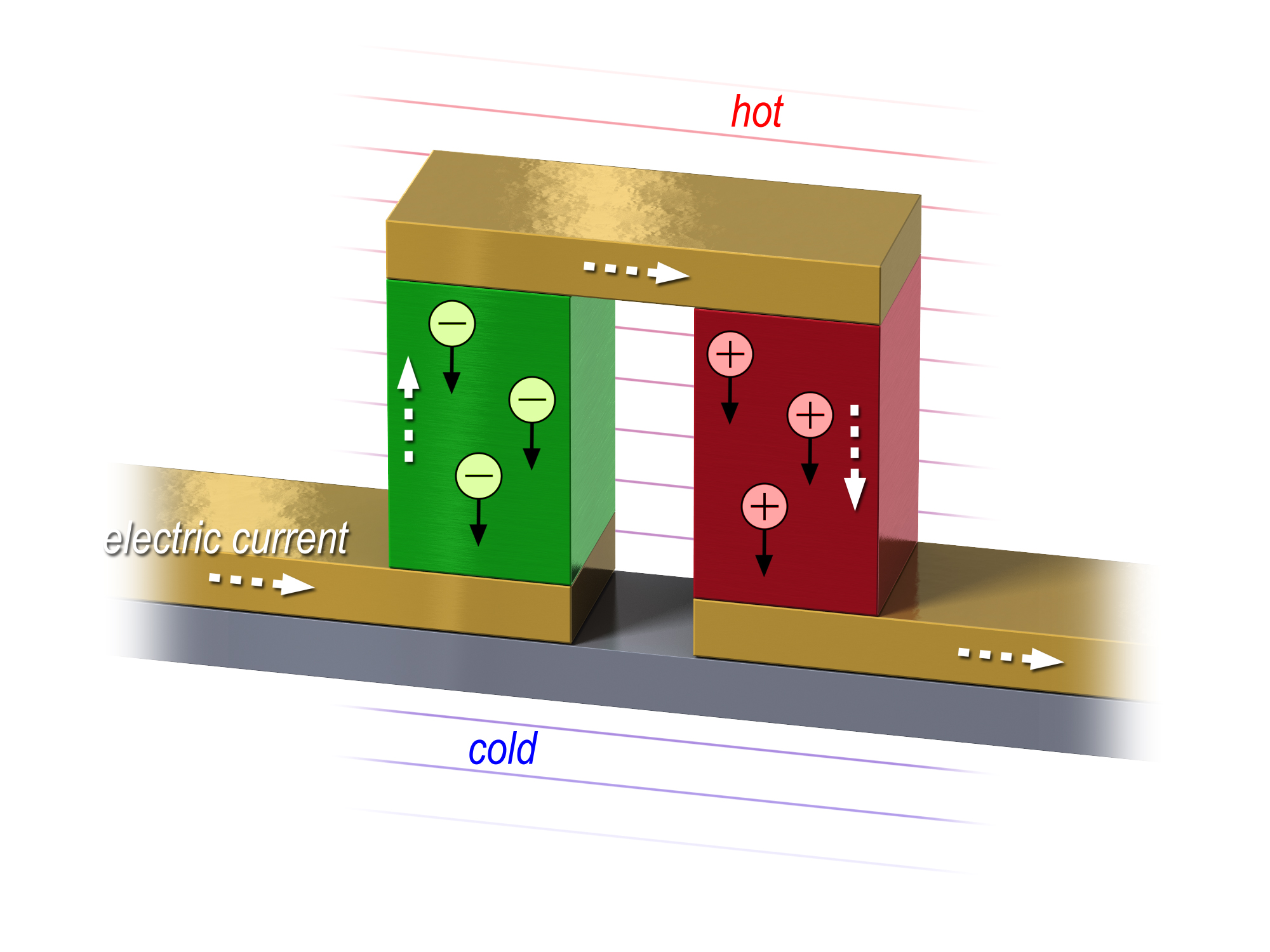}
	\caption{{\bf Traditional thermocouple.} A traditional thermocouple device consists of a $n$-type material (in green) and a $p$-type material (in red) connected electrically in series, but parallel with respect to the temperature gradient. The charge carriers move from the hot end to the cold end (black arrows), generating an electrical current (white dashed arrows).}
	\label{Fig1}
\end{center}
\end{figure}

A typical thermoelectric device consists of both $p$-type and $n$-type materials, as shown in Fig \ref{Fig1}. The key to realize an efficient thermoelectric device is finding materials with high thermoelectric performance, commonly expressed in terms of the dimensionless \emph{figure of merit}~\cite{Ioffe1957, Snyder2007}
\begin{equation}
	zT = \frac{\alpha_e^2 / L}{1 + \frac{\kappa_\ph}{\kappa_e}}
\end{equation}
where $\alpha_e$, $\kappa_\ph$, $\kappa_e$ and $L$ are the Seebeck coefficient, the phonon thermal conductivity, the thermal conductivity of the electrons and the Lorenz number, respectively. Following the Wiedemann-Franz law, the Lorenz number $L$ is defined by the ratio \cite{AshcroftMermin}
\begin{equation}\label{Lorenz number}
	L = \frac{\kappa_e}{\sigma_eT},
\end{equation}
where $\sigma_e$ is the electrical conductivity and $T$ is the temperature. The theoretical value for the Lorenz number of degenerate metals is given by $L_{0}={\pi^2\over3}({k_B\over e})^2 = 2.44 \times 10^{-8}$ V$^2$ K$^{-2}$ \cite{Sommerfeld1927, Sommerfeld1928}. Indeed, experimentally measured values of $L$ show only small deviations from this Sommerfeld value, namely $L=1.7-2.6\times 10^{-8}$ V$^2$ K$^{-2}$ in elemental metals, metallic compounds and semiconductors \cite{Lorenz1881a, Lorenz1881b,Kaye1966, Kumar1993}.
	
\begin{figure}
\begin{center}
	\includegraphics[width=\columnwidth]{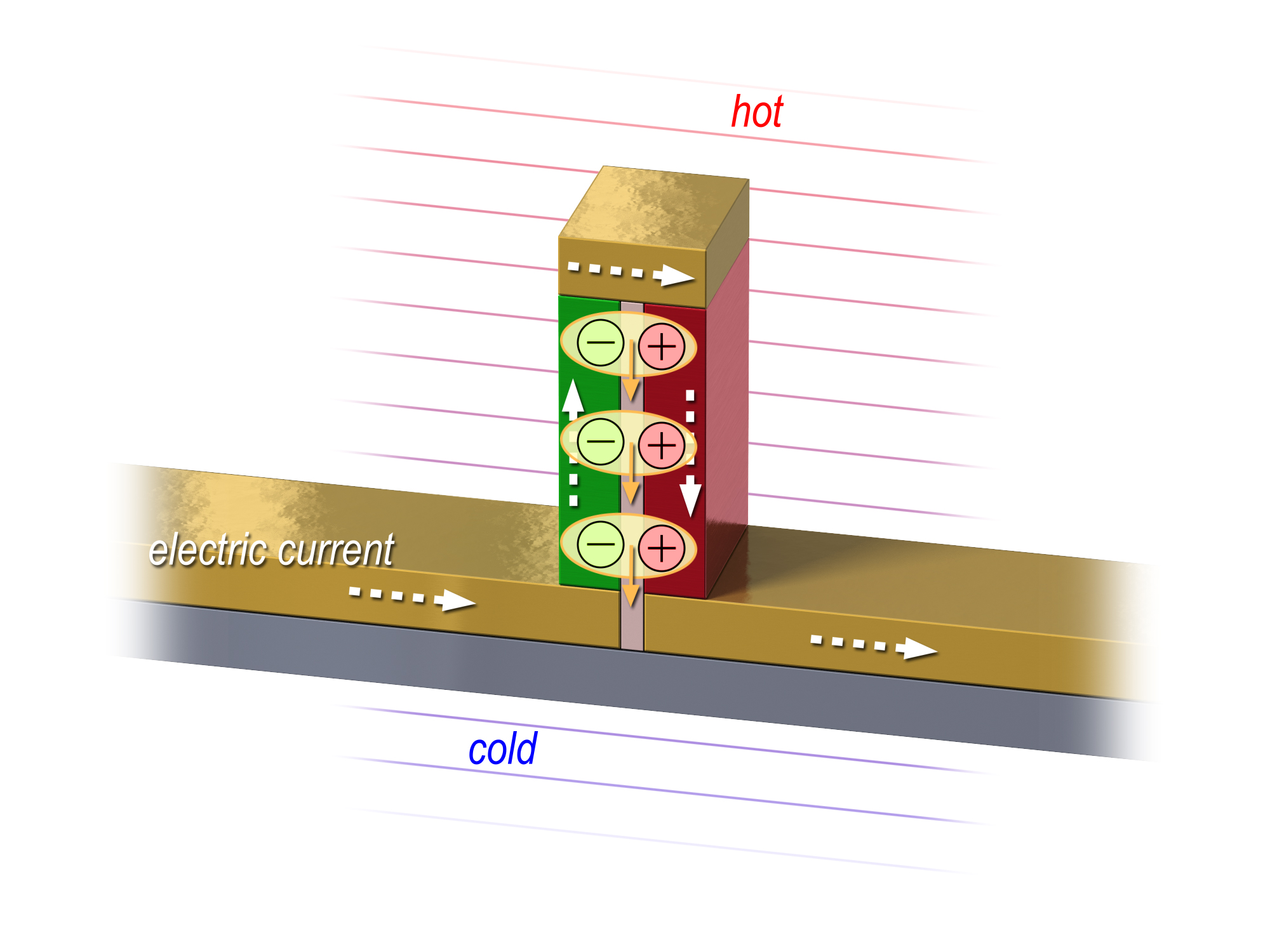}
	\caption{{\bf Bilayer exciton thermocouple.} In a bilayer exciton thermocouple, as presented in this paper, the $p$- and $n$-type materials are brought close together so that bilayer excitons can form. In the thermoelectric device, the excitons move from the hot side to the cold side (orange arrows) generating an electric counterflow (white dashed arrows). Due to the bosonic nature of the excitons, the exciton thermocouple can have a higher figure of merit than the traditional thermocouple of Fig. \ref{Fig1}.}
	\label{Fig2}
\end{center}
\end{figure}

Therefore, the obvious strategy to maximize the thermoelectric figure of merit is by increasing the thermopower $\alpha$ and decreasing the phonon thermal conductivity $\kappa_\ph$. Indeed, the central maxim of most thermoelectric research is summarized by the phrase "phonon glass - electron crystal" \cite{Slack,DiSalvo1999,Takabatake}. A different route to increase $zT$, however, is to involve low-dimensional nanostructures \cite{Hicks1993,Giazotto}. Specifically, the use of 2d quantum wells \cite{Hicks1996}  has enhanced the bulk value of $zT$ for the "state-of-the-art" hole-doped semiconductor Bi$_2$Te$_3$ from its bulk value of $zT=1$ \cite{Kuznetsov2002} to $zT \sim 2.4$ for thin films \cite{Venkatasubramanian2001}. In a 2d electron gas of doped SrTiO$_3$ a $zT\sim 2.4$ has been achieved \cite{Ohta2007}. However, to overcome the competition of conventional heat pumps regarding efficiency we need to find new materials to enhance $zT$ to 4 or greater \cite{Vining2009,DiSalvo1999}.  In this Article we present a concept that has the potential to increase of the figure of merit $zT$ by a factor of 60. For this we need to mobilize \emph{bilayer excitons}.

An exciton is the bound state of an electron and a hole. In a bulk material the electron and hole can recombine and thus annihilate the exciton. In order to avoid this annihilation, it has been proposed to spatially separate the electrons and holes in two different layers \cite{Shevchenko:1976p4950,Lozovik:1976p4951}. Such a heterostructure of a $p$-type and $n$-type material can support bilayer excitons. Since the bound state of an electron and a hole is effectively a boson, it can Bose condense, and substantial experimental and theoretical efforts have been devoted to the  realization of a bilayer exciton condensate \cite{Moskalenko:2000p4767, Eisenstein:2004p4770, High:2012p5349,Rademaker,Rademaker2, Rademaker3}.

The bilayer exciton system displays a remarkable similarity to a thermoelectric couple, as shown in Fig. \ref{Fig1}. Such a device consists of a $p$-type and $n$-type thermoelectric material connected in series \cite{Snyder2007}. A heat gradient will cause the charge carriers to diffuse from the hot side to the cold side, thus converting the heat gradient into an electrical current. Now imagine that one brings the $p$- and $n$-type material close to each other. At distances of order nanometers the Coulomb attraction between the electrons and holes can bind them together into bilayer excitons. As shown in Fig. \ref{Fig2}, the application of a heat gradient will cause these excitons to flow from the hot side to the cold side. Even though excitons are charge neutral, the fact that the holes and electrons are spatially separated implies that an exciton current amounts to two opposite currents in the two layers \cite{SuMacDonald}. If one closes the circuit in the counterflow set-up, see Fig. \ref{Fig2}, the exciton current is converted into an electrical current. Such excitonic counterflow conductivity has been demonstrated experimentally \cite{Finck:2011p5284}.

Though most bilayer exciton research focusses on the Bose condensation of excitons, for thermoelectric applications uncondensed excitons are required. Although the counterflow conductivity of the exciton Bose condensate is infinite due to the superfluidity, the Seebeck coefficient vanishes faster and the figure of merit drops to zero \cite{Faniel2005}. However, we will show here that the high-temperature properties of a gas of bilayer excitons will lead to an exceptionally large thermopower, combined with a small Lorenz number. In addition, the remarkably weak coupling between excitons and phonons, the dominant scattering mechanism at room temperature, enhance the electric counterflow conductivity and gives rise to a high exciton thermal conductivity which reduces the effect of the phonon thermal conductivity $\kappa_\ph$.

These are the two proposed pillars of bilayer exciton thermoelectrics: better thermopower, and better conductivities. We suggest two possible material choices for the bilayer exciton thermocouple: a Bi-based system of $p$- and $n$-doped Bi$_2$Te$_3$, and an oxide system consisting of $n$-doped SrTiO$_3$ and $p$-type Ca$_3$Co$_4$O$_9$. The expected enhancement of thermopower and thermal conductivities are shown in Table \ref{TheTable}, which comprises our main result. In the remainder of this Article we first derive the relations between the exciton correlation functions and the counterflow transport coefficients, in Sec.~\ref{Sec1}. Subsequently, we will describe the enhancement of the thermopower in Sec.~\ref{SecThermo} and the conductivity in Sec.~\ref{SecCond}. We end in Sec.~\ref{SecExp} with suggestions towards experimental realization of a bilayer exciton thermocouple.

\begin{table}
\label{Tb:Exciton1}
\begin{ruledtabular}
\centering
\begin{tabular}{p{2cm}ccccccc}
    \centering{\bf materials} & $n$ &  $g_sg_0$ & $\alpha$ & $\kappa_\ph $ & $\kappa_e(\kappa_{ex})$& $ zT $ \\
     \hline 
     \centering{doped Bi$_2$Te$_3$ } &    $10^{19}$      &     $2\times6$       &   220       & 0.8    & 0.8 &   1.0     \\
     
     \centering\bf{Bi-based Exciton}   &  \tabincell{c}{\\$\mathbf{3\times 10^{18}}$}  &    \tabincell{c}{\\$\mathbf{4\times36}$}     &   \tabincell{c}{\\$\mathbf{1060}$}  & \tabincell{c}{\\$\mathbf{1.6}$} & \tabincell{c}{\\$\mathbf{7.0}$ }&   \tabincell{c}{\\$\mathbf{63}$}\\
     \hline 
     \centering{Nb-SrTiO$_3$}    &    $6\times10^{20}$      &    $2\times3$       &   150          & $8.2$ &$0.9$  & 0.08  \\
     
    \centering{Ca$_3$Co$_4$O$_9$} &  $1.9\times10^{21}$     &      $2$            &   120          & 3.0 &0.5    &  0.08 \\
    
    \centering\bf{O-based Exciton}&  \tabincell{c}{\\$\mathbf{1.0\times10^{21}}$} & \tabincell{c}{\\$\mathbf{4\times3}$}  &   \tabincell{c}{\\$\mathbf{350}$} &  \tabincell{c}{\\$\mathbf{11.2}$}        & \tabincell{c}{\\$\mathbf{8.3}$}   &  \tabincell{c}{\\$\mathbf{3.6}$} \\
    
\end{tabular}
\end{ruledtabular}
\caption{{\bf Room temperature properties of traditional thermoelectrics compared to bilayer exciton systems.} Here we compare three traditional thermoelectrics (doped Bi$_2$Te$_3$ \cite{Snyder2007,Kuznetsov2002}, Nb-doped SrTiO$_3$ \cite{OxideReview1,OxideReview2,OxideReview3} and Ca$_3$Co$_4$O$_9$ \cite{OxideReview1,OxideReview2,OxideReview3}) with two bilayer exciton thermocouples based on known experimental data. The Bi-based exciton bilayer consists of $p$ and $n$-doped Bi$_2$Te$_3$, the oxide-based exciton bilayer is a SrTiO$_3$/Ca$_3$Co$_4$O$_9$ heterostructure.
The columns display the optimal carrier density $n$(cm$^{-3}$), degeneracies $g_sg_0$, thermopower $\alpha$($\mu $VK$^{-1}$), phonon thermal conductivity $\kappa_\ph$(W m$^{-1}$K$^{-1}$), electron/exciton thermal conductivity $\kappa_e(\kappa_{ex})$ (W m$^{-1}$K$^{-1}$) and the figure of merit $zT$. For bilayer excitons, the $\kappa_\ph$ is taking into account the contribution of phonons in both layers.
Clearly, the exciton systems exhibit a tremendous enhancement of the thermoelectric figure of merit $zT$.}
\label{TheTable}
\end{table}

\section{Exciton transport and counterflow coefficients}
\label{Sec1}
Since an exciton is the bound state of an electron and a hole, naively one would expect that excitons would not display any electrical effects. However, in the counterflow set-up the spatial separation of the electrons and holes of the bilayer exciton allow for an electric response. In this section we will discuss how bilayer exciton response functions can be used to express the figure of merit for a bilayer exciton thermocouple.

\begin{figure}
	\includegraphics[width=\columnwidth]{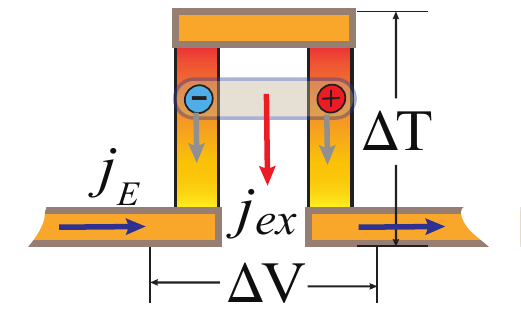}
	\caption{{\bf Thermal and electrical response of the excitons.} In the counterflow construction both electrons and holes will perceive the same electrical forces (grey arrows) when there is a voltage different between the external leads. The exciton particle current $j_{ex}$(red arrow) is driven by both a temperature gradient $\Delta T$ and a voltage $\Delta V$, and generates an electrical current $J_E$(blue arrows) which is proportional to the exciton particle current $J_E=e j_{ex}$.}
	\label{Fig4}
\end{figure}

We consider the limit where the Coulomb attraction between electrons and holes is strong enough to allow us to treat the exciton as a composite particle moving in a 2d layer. The electrons and holes are spatially separated in the direction perpendicular to this layer. In this limit, we envisage a voltage and a temperature gradient as displayed in Fig.~\ref{Fig4}. Denoting the exciton particle current and heat current as $J_{ex}$ and $J_{exQ}$, respectively, the traditional Onsager relations for this system are
\begin{eqnarray}
j_\ex&=&-\frac{1}{T}L^{(11)}\Delta\bar{\mu}+L^{(12)}\Delta(\frac{1}{T})\nonumber\\
     j_{\ex Q}&=&-\frac{1}{T}L^{(21)}\Delta\bar{\mu}+L^{(22)}\Delta(\frac{1}{T})
\end{eqnarray}
where $\Delta\bar{\mu}=\Delta\mu+e\Delta V/L$ is the effective chemical potential gradient due to the voltage $\Delta V$ between the external leads. Note that due to this voltage, both electrons and holes perceive an electric force $e\Delta V/2L$ in the same direction. Thus, effectively the force acting on an exciton equals $e\Delta V/L$.

All of the $L^{ab}$ coefficients are directly associated with different exciton current-current correlation functions,
\begin{eqnarray}
L^{(11)}&=&-\frac{1}{\beta}\int_0^\infty dt e^{-st}\int_0^\beta d\beta' \Tr[\rho_0 j_{ex}(-t-i\beta')j_{ex}]\nonumber\\
L^{(12)}&=&-\frac{1}{\beta}\int_0^\infty dt e^{-st}\int_0^\beta d\beta' \Tr[\rho_0 j_{ex}(-t-i\beta')j_{exQ}]\nonumber\\
L^{(22)}&=&-\frac{1}{\beta}\int_0^\infty dt e^{-st}\int_0^\beta d\beta' \Tr[\rho_0 j_{exQ}(-t-i\beta')j_{exQ}]\nonumber \\ \label{correlation function}
\end{eqnarray}
Finally, when one exciton moves from the hot side to the cold side, there is effectively one unit of charge moved from the left to the right side lead. This implies that the net electrical current $J_E$ between the leads generated by the exciton particle current is given by $J_E=e j_{ex}$. Using this insight, we are now in the position to derive all the relevant transport coefficients. For example, the exciton counterflow dc conductivity $\sigma_{\ex}$ can be found by setting $\Delta\mu=\Delta T=0$ in the Onsager relations, which leads to
 \begin{eqnarray}
	J_E&=&\sigma_{\ex} E\nonumber\\
	\sigma_{\ex}&=&\frac{e^2}{T}L^{(11)}\label{exconductivity}
\end{eqnarray}
Similarly, the thermopower and thermal conductivity can be written as:
\begin{eqnarray}
    \kappa_{\ex}&=&\frac{1}{T^2}[L^{(22)}-\frac{(L^{(12)})^2}{L^{(11)}}] \label{exthermal}\\
    \alpha_{\ex}&=&\frac{1}{eT}\frac{L^{(12)}}{L^{(11)}}\label{exthermopower}
\end{eqnarray}
From the electrical conductivity \eqref{exconductivity} and the thermopower \eqref{exthermopower} of the whole device, we see that when we consider the counterflow transport coefficients in the limit of strongly bound bilayer excitons, it is equivalent to treat {\it excitons as bosonic particles with charge $e$ and mass $2m$}. Consequently, for the whole exciton system the figure of merit of the counterflow set-up can be written as:
\begin{eqnarray}
     zT&=&\frac{\sigma_{\ex}\alpha_{\ex}^2 T}{\kappa_{\ex}+2\kappa_{ph}}
\end{eqnarray}
where $2\kappa_{ph}$ is thermal conductivity contribution from phonons in both layers.  

In the remainder of this Article, we consider temperatures much higher than the critical temperature $T_c$ of the exciton condensate, but lower than the exciton binding energy. Additionally, we only consider the case in which the exciton density is low, {it i.e.} the dilute limit, to maximize the thermopower. Thus, within these two limits, the bilayer exciton system approaches the classical limit of bosonic particles and we can easily apply the Drude theory for all transport coefficients. This will be done in the next two sections.

\section{Enhanced thermopower}
\label{SecThermo}
A dominant factor that determines the figure of merit is the thermopower. For Fermi liquid materials such as metals or degenerate semiconductors, the thermopower Seebeck coefficient $\alpha$ is given by \cite{AshcroftMermin}
\begin{equation}
|\alpha| = \frac{\pi^2}{3} \frac{k_B}{e} \left( \frac{T}{T_F} \right),
\end{equation}
where $k_B$ is the Boltzmann constant, $e$ is the elementary electric charge and $T_F$ is the Fermi temperature of the material. Therefore, a natural way to increase the thermopower is to reduce the charge carrier density $n$, since $T_F \sim n^{2/3}$.

Bilayer excitons, on the other hand, behave like hard-core bosons on a lattice. Following Eqn. \eqref{exthermopower}, the counterflow thermopower can be written as
\begin{equation}\label{exthermopower2}
\alpha_{\ex}=\frac{1}{eT}\frac{S^{(12)}}{S^{(11)}}-\frac{\mu}{eT}
\end{equation}
where $S^{(11)}=L^{(11)}$ and $S^{(12)}=L^{(12)}-\mu L^{(11)}$ are the exciton correlation functions between the energy current and the exciton particle current, similar to Eqn. \eqref{correlation function}. For temperatures well above the condensation temperature $T_c$ and in the dilute limit, the contribution from the correlation function in \eqref{exthermopower2} is minor and the excitons behave as classical particles in which case the thermopower is given by the entropy transported per particle, divided by the transported charge per particle \cite{Beni,ChaikinBeni}.

Recall that when an exciton is created at the top of the counterflow construction of Fig. \ref{Fig2} and is moved through the bilayer to the bottom, an excess negative charge has been created in the left lead and an excess positive charge in the right lead. Effectively, this implies that a single electric charge has been moved from the left lead to the right lead. Therefore, one moving exciton in the counterflow set-up transports one unit of charge.

The entropy of a bilayer exciton is given by the logarithm of the number of possible configurations, whereby spin degeneracy $g_s$ and orbital degeneracy $g_o$ play an important role \cite{Koshibae2001,Wang2003}. For excitons on a crystalline lattice with density $\rho$ per unit cell, we find that the excitonic thermopower is given by
\begin{equation}
	\alpha_\ex = \frac{k_B}{e} \left( \ln g_s g_o + \ln \frac{1-\rho}{\rho} \right).
	\label{ExcitonThermo}
\end{equation}
The exciton spin degeneracy is usually $g_s=2 \times 2=4$ since both the electron and hole can have two spin states, and the singlet-triplet splitting is negligible compared to the temperature. Similarly, the exciton orbital degeneracy $g_0$ is the product of the electron and hole orbital degeneracy, {\it e.g.} $6\times6$ for doped Bi$_2$Te$_3$.

In Table \ref{TheTable} we compare the Seebeck coefficient of traditional thermoelectric materials to bilayer exciton systems. The entropic increase of the excitons, as a result of the orbital and spin degeneracies, gives rise to a large thermopower for the bilayer exciton thermocouple.

Recall that for most free electron system, the Lorenz number was found to be constant \cite{AshcroftMermin}. However, in the case of excitons, we no longer have free fermionic electrons but instead we are dealing effectively with bosonic particles of charge $e$. At high temperatures, the Lorenz number for dilute bosons is given by \cite{Alexandrov1993}
\begin{equation}
	L_\ex = 2 \left( \frac{k_B}{e} \right)^2,
\end{equation}
which is smaller than the Wiedemann-Franz value by a factor $\frac{L_{\mathrm{WF}}}{L_\ex} \approx 1.6$.

\section{Enhanced conductivity}
\label{SecCond}
Now we discuss the thermal conductivity of excitons. It is known that bulk excitons have an exceptionally large mobility. For example, in cuprous oxide Cu$_2$O at low temperatures the exciton mobility is found to be $\mu_\ex \sim 10^7$ cm$^2 / $Vs \cite{Trauernicht}, orders of magnitude larger than the mobility of charged carriers in this compound \cite{Lee2011}. Measurements on low temperature bilayer exciton mobility in GaAs/AlGaAs double quantum wells show similarly high mobilities \cite{Voros2005}. Recent results at $T=100$ K in GaAs/AlGaAs double quantum wells suggest that the high mobility of excitons persists up to higher temperatures \cite{Grosso2009}.

Following the sparse experimental results, we can make a qualitative prediction for the room temperature exciton counterflow conductivity. For any kind of particle, its resistance depends on impurity scattering, scattering with phonons and particle-particle scattering. In a pure dilute sample at room temperatures, the dominant contribution to the resistivity is given by phonon-scattering.

In the case of charged carriers, the scattering with acoustic phonons depends on the deformation potential $D$. Even though excitons are charge-neutral, the spatial separation of the electrons and holes still allows for a coupling between the acoustic phonons and the exciton dipole moment. The leading order long-wavelength matrix element between excitons and acoustic phonons is given by a combination of the electron and hole contribution \cite{Moskalenko:2000p4767,Toyozawa}
\begin{equation}
	\sqrt{\frac{2\hbar k}{M u}} \left( D_e  - D_h \right)
	\label{MatrixExPh}
\end{equation}
where $k$ is the phonon wavevector, $u$ is the speed of sound, $M$ is the ion mass, and $D_{e,h}$ are the deformation potentials of the electron and hole, respectively \cite{Moskalenko:2000p4767,Toyozawa}. The minus sign in Eqn. (\ref{MatrixExPh}) is due to the opposite charge of the electron and hole, and thus leads to a relatively small exciton-phonon coupling when $D_e \approx D_h$. Note that Eqn. (\ref{MatrixExPh}) is valid only for small distances between the electrons and holes, as for example experienced in bulk Cu$_2$O. There the radius of a bulk exciton is approximately $7 \AA$, hence as long as the interlayer excitons are of similar size we expect the above model to hold.

Thus, if one is able to construct a heterostructure such that the deformation potentials $D_{e,h}$ for the electrons and holes are almost equal, one finds an extremely low exciton-phonon scattering rate. In fact, relative to the electron-phonon scattering rate,
\begin{equation}
	\frac{\tau_{\ex-\mathrm{ph}}}{\tau_{e-\mathrm{ph}}}
	\sim \left( \frac{D}{\Delta D} \right)^2
\end{equation}
where $\Delta D$ is the absolute difference between the electron deformation potential and the hole deformation potential. From the scattering rate $\tau$, the electric counterflow conductivity is given by $\sigma_{\ex}= ne^2 \tau_{\ex}/2m_\ex$ with $m_\ex\approx2m_e$. Using the Wiedemann-Franz law we then find the change in thermal conductivity of the excitons with respect to the electronic thermal conductivity,
\begin{align}
    \frac{\sigma_{\ex}}{\sigma_e}&=\frac{\tau_{\ex-\mathrm{ph}}}{\tau_{e-\mathrm{ph}}}
		\frac{m_e}{m_{\ex}}
		\sim \frac{1}{2}\times \left( \frac{D}{\Delta D} \right)^2,\nonumber\\
	\frac{\kappa_{\ex}}{\kappa_e}
		&=\frac{\sigma_{\ex}}{\sigma_e}
		\frac{L_{\ex}}{L_{WF}}
		\sim
		0.3 \times \left( \frac{D}{\Delta D} \right)^2.
\end{align}
The aforementioned orders of magnitude increase of mobility in bulk Cu$_2$O \cite{Trauernicht} suggests that a $10 \%$ difference between the deformation potentials of the electrons and holes is a reasonable assumption. This will lead to a factor fifty better electric conductivity, and a factor thirty better thermal conductivity of the excitons. We used this estimate of the deformation potentials to obtain the results of Table \ref{TheTable}.

\section{Experimental realization}
\label{SecExp}
An idealized exciton thermocouple could therefore have a $zT$ of more than 60 based on current experimental parameters, see Table \ref{TheTable}. This is, however, the optimistic theorist perspective. In reality, many pitfalls and engineering problems will lower the figure of merit. Nonetheless, with the ideal $zT \sim60$ there is much room for shortfalls to still arrive at a sizable figure of merit.

The biggest experimental challenge is to actually make bilayer excitons without condensing them, just like it is difficult to get uncondensed electron-electron pairs. Most material proposals for bilayers exciton condensation, such as semiconductor quantum wells \cite{High:2012p5349}, double layer graphene \cite{Lozovik:2008p4877,Zhang:2008p4895,Dillenschneider:2008p4896,Min:2008p4795,Kharitonov:2008p5044} or topological insulators \cite{Seradjeh:2009p4980} function in the BCS limit, which means that bilayer excitons are not bound at temperatures above the condensation temperature. Instead, we need materials with strong exciton binding, which can be achieved by bringing the layers close to each other (of the order of several unit cells), to minimize the electron and hole kinetic energy and to ensure that the electrons and holes have similar effective mass and deformation potential. For this end we propose two possible material choices for the bilayer exciton thermocouple: bismuth-based bilayers and oxide-based bilayers.

\begin{figure}
\begin{center}
	\includegraphics[width=\columnwidth]{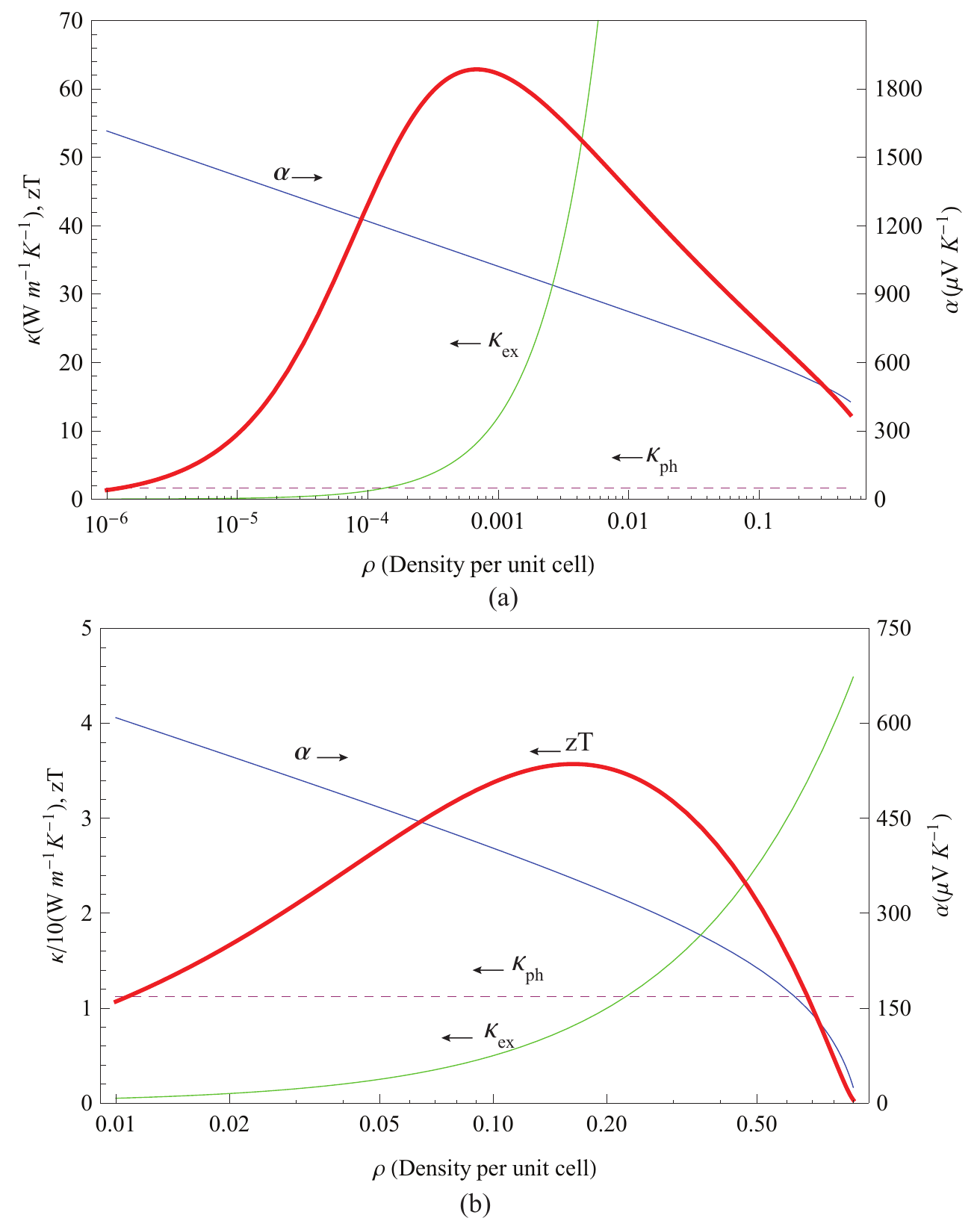}
	\caption{{\bf Density dependence of the thermoelectric properties.}
	{\bf a.} Here we show the figure of merit $zT$, the exciton thermopower $\alpha_\ex$, the phonon thermal conductivity $\kappa_\ph$ and the exciton thermal conductivity $\kappa_\ex$ for the $p$- and $n$-doped Bi$_2$Te$_3$  bilayer. The optimal value of $zT$ is used in Table \ref{TheTable}.
	{\bf b.} Same as {\bf a}, but for the bilayer of $n$-doped SrTiO$_3$ and $p$-type Ca$_3$Co$_4$O$_9$.}
	\label{Fig3}
\end{center}
\end{figure}

Oxide heterostructures \cite{Ribeiro2006,Millis:2010p5231,Rademaker,Rademaker2,Rademaker3}, that themselves  show good thermoelectric properties \cite{OxideReview1,OxideReview2,OxideReview3}, will show a significant enhancement of the figure of merit. Many oxide materials have a layered lattice structure, such as $n$-doped SrTiO$_3$ \cite{Ohta2005,Ohta2007}, and the $p$-type cobaltates Ca$_3$Co$_4$O$_9$ and Na$_x$CoO$_2$ \cite{OxideReview2}. Techniques such as molecular beam epitaxy (MBE) and pulsed laser deposition (PLD) make it currently possible to grow heterostructures of layered oxide materials with unit-cell precision. We propose to construct a bilayer system with a few unit cell layers of Nb-doped SrTiO$_3$, followed by about 6 unit cell layers of an oxide insulator like SrTiO$_3$ to avoid recombination of electrons and holes, and then topped a few unit cell layers of the $p$-type Ca$_3$Co$_4$O$_9$. This construction should be repeated in a superlattice structure, with insulating materials between each bilayer, to obtain a material with a bulk thermoelectric effect. Using ion-beam etching ramp-edges can be made, so that one can separately contact each $n$-type and $p$-type layer for the deposition of gold contacts.

We have chosen these materials because they can be fabricated with similar charge carrier densities of $n \sim 10^{20}$ cm$^{-3}$, and the carriers have similar room temperature electronic thermal conductivities $\kappa_e \sim 0.5$ W m$^{-1}$ K$^{-1}$ \cite{Ohta2005,Shikano2003,Luo2006}. SrTiO$_3$ has 3-fold orbital degeneracy, while the cobaltate is a one-band material. Together with the spin entropy, we expect that at the given densities the exciton thermopower is $\alpha_\ex \sim 650$ $\mu$V K$^{-1}$, which is even a conservative estimate given the $\alpha \sim 1000$ $\mu$V K$^{-1}$ results in SrTiO$_3$ superlattices \cite{Ohta2007}. From our earlier analysis we expect that the exciton thermal conductivity is greatly enhanced. Since the phonon thermal conductivity in oxides is usually the limiting factor in obtaining a high $zT$, the large exciton thermal conductivity yields several orders of magnitude enhancement of $zT$, see Table \ref{TheTable}. The density-dependence of $\alpha$, $\kappa_\ex$ and $zT$ for the oxides is displayed in Fig. \ref{Fig3}a, showing that for a wide window of exciton densities the SrTiO$_3$/Ca$_3$Co$_4$O$_9$ bilayer system has a sizable figure of merit.

The other possibility is to form heterostructures of $p$- and $n$-doped Bi$_2$Te$_3$ \cite{Kuznetsov2002}, separated at a distance of $\sim 2$-$3$ nm.
Molecular beam epitaxy (MBE) and chemical vapor deposition (CVD) have been used extensively to fabricate nanostructures of bismuthtellurides \cite{Venkatasubramanian2001}. The fact that we have the same material used for both the $p$-type and $n$-type layer implies that the exciton binding and the exciton-phonon coupling are optimal. Together with the sixfold band degeneracy of Bi$_2$Te$_3$ we find an extraordinary large Seebeck coefficient, $\alpha \sim 1115$ $\mu$V K$^{-1}$. We thus arrive at an extremely large $zT \sim 60$, see Table \ref{TheTable}. The density-dependence of the thermopower and the thermal conductivity is shown in Fig. \ref{Fig3}b.

The main question is of course whether excitons would form at room temperature. If we take the insulating layer to have a dielectric constant of about $\epsilon \sim 20$, then at a distance of 2.3 nm the electron-hole attraction is $V \sim 30$ meV $=350$ K. Given this estimate of the binding energy, most of the electrons and holes will be bound into bilayer excitons at room temperature.

Nonetheless, at room temperature it is likely that bilayer excitons coexist with unbound electrons and holes. Naturally, this will reduce the figure of merit, as some of the currents are carried by the less-efficient electrons and holes. A further experimental challenge is therefore to dilute the number of unbound electrons and holes, for example by bringing the $p$ and $n$-type materials closer together or to change the insulator layer.

Note that in our current proposal we have kept the phonon thermal conductivity constant at its relatively large bulk value, emphasizing only the desired enhancement due to the excitonic effects. This implies that a further enhancement of the figure of merit can be achieved by combining exciton physics and one of the known techniques to reduce phonon thermal conductivity.

In conclusion, we have shown qualitatively that the thermoelectric efficiency of a bilayer exciton system can be significantly enhanced. We suggest that the nanotechnological engineering of thermoelectric devices should therefore aim at using bilayer excitons, instead of electrons and holes, to generate electricity out of heat.

\emph{Acknowledgements} - We thank Binghai Yan, John Mydosh, Hans Hilgenkamp, Zhi-Xun Shen, Yayu Wang, Jim Eisenstein, Allan MacDonald and Aaron Finck for discussions, and Jeroen Huijben (Nymus3D) for designing the figures. LR is supported by the Netherlands Organisation for Scientific Research (NWO) via a Rubicon grant.

\end{document}